# Accuracy Limits of Polarization-Independent Optical Depolarizers Based on Rotating Waveplates

Reinhold Noé[(1,2)], Benjamin Koch[(1,2)]
(1) Paderborn University, EIM-E, Warburger Str. 100, D-33098 Paderborn, Germany, e-mail: noe@upb.de
(2) Novoptel GmbH, Helmerner Weg 2, D-33100 Paderborn, Germany, E-mail: info@novoptel.com

*Abstract*—Optical depolarizers for monochromatic waves which work independent of input polarization can be built from cascaded electrooptic rotating waveplates. If the waveplate retardations deviate from their desired values then the worst-case residual degree-of-polarization DOPmax is larger than its desired value 0. In a depolarizer consisting of one rotating halfwave and one rotating quarterwave plate, DOPmax roughly equals the retardation error, which is <<1. However, with just one rotating quarterwave plate more, DOPmax roughly equals the square of the retardation error which is a much smaller value. Thereby depolarizer accuracy is substantially improved. Waveplate sequence and rotation frequency combinations suitable for fast depolarization are discussed.

## I. Introduction

Optical depolarizers serve to make measurements or transmission independent of polarization, typically independent of polarization-dependent loss (or gain) in optical fibers. For polychromatic signals, Lyot depolarizers and similar highly dispersive devices work well. Monochromatic signals, or such signals which do not tolerate added polarization mode dispersion, require time-variable depolarizers which usually contain time-variable retarders. Examples are given in [1–3].

Depolarizers are usually required to work independent of input polarization, in order to make them usable for any input signal. Cascaded TE-TM phase shifters and TE-TM mode converters can achieve this, or rotating quarterwave (QWP) and halfwave plates (HWP). It is near at hand to implement electrooptic waveplates in X-cut, Z-propagation LiNbO$_3$ [4–6]. Such polarization-independent depolarizers were of interest in the early times of optically amplified trunk lines with just one single-polarization signal. They could suppress polarization holeburning of EDFAs.

In [7], the acoustooptic equivalent of rotating QWP and HWP was realized. The difference between two electrical frequencies, used to drive acoustooptic mode converters, was just 250 kHz, corresponding to a depolarization interval $T$ or period of 4 μs. Unless waveguides are made longitudinally inhomogeneous, depolarization interval could not be reduced. More importantly, function depended strongly on wavelength and temperature, and the whole device introduced severe PMD with a DGD of about 20 ps.

Today, polarization-independent depolarizers are of high interest in fiberoptic communication. They are requested for control or supervisory signals, over very long distances of terrestrial and submarine lines. But the depolarization interval $T$ is required to be < 1 μs for such applications. Function should of course not strongly depend on wavelength and temperature. Due to the above, electrooptic rotating waveplates on a low-loss X-cut, Z-propagation LiNbO$_3$ integrated photonic component appear to be the best choice.

The purpose of this work is to detail what must be done to minimize the maximum residual degree-of-polarization $DOP_{max}$ even if the rotating waveplates are nonideal, namely if they have retardations which differ from the desired ones.

## II. Depolarizer with 2 rotating waveplates

Let $\mathbf{R}$ be the 3×3 rotation matrix of a depolarizer in the normalized Stokes space. Its normalized output Stokes vector $\mathbf{S}_{out}$ is calculated from the normalized input Stokes vector $\mathbf{S}_{in}$ as

$$\mathbf{S}_{out} = \mathbf{R}\mathbf{S}_{in}. \qquad (1)$$

The length of the time-averaged $\mathbf{S}_{out}$ is the degree-of-polarization

$$DOP = \left|\overline{\mathbf{S}}_{out}\right| = \left|\overline{\mathbf{R}}\mathbf{S}_{in}\right|. \qquad (2)$$

The time-averaged rotation matrix must be the zero matrix $\overline{\mathbf{R}} = \mathbf{0}$ in order to achieve a $DOP = 0$ for all input polarizations $\mathbf{S}_{in}$. In practice, the largest singular value of $\overline{\mathbf{R}}$, equal to the square root of the largest eigenvalue of $\overline{\mathbf{R}}^T \overline{\mathbf{R}}$ ($T$ = transpose), gives the largest $DOP_{max}$ that must be tolerated.

A waveplate with linearly polarized eigenmodes is characterized by

$$\mathbf{R}(\delta,\psi) = \begin{bmatrix} (1+\cos\delta)/2 + \cos 2\psi(1-\cos\delta)/2 & \sin 2\psi(1-\cos\delta)/2 & \sin\psi\sin\delta \\ \sin 2\psi(1-\cos\delta)/2 & (1+\cos\delta)/2 - \cos 2\psi(1-\cos\delta)/2 & -\cos\psi\sin\delta \\ -\sin\psi\sin\delta & \cos\psi\sin\delta & \cos\delta \end{bmatrix} \qquad (3)$$

with retardation $\delta$ and eigenmode orientation angle $\psi$ on the equator of the Poincaré sphere.

In the following we assume waveplates to have small retardation errors $\xi$ with $|\xi| << 1$, $\sin\xi \approx \xi$, $\cos\xi \approx 1 - \xi^2/2 \approx 1$.

A nonideal quarterwave plate (QWP) with retardation $\delta = \pi/2 + \xi$ then has the (approximate) rotation matrix

$$\mathbf{QWP}(\xi,\psi)$$
$$\approx \begin{bmatrix} (1-\xi)/2 + \cos 2\psi(1+\xi)/2 & \sin 2\psi(1+\xi)/2 & \sin\psi \\ \sin 2\psi(1+\xi)/2 & (1-\xi)/2 - \cos 2\psi(1+\xi)/2 & -\cos\psi \\ -\sin\psi & \cos\psi & -\xi \end{bmatrix}. \quad (4)$$

A nonideal halfwave plate (HWP) with retardation $\delta = \pi + \xi$ has the (approximate) rotation matrix

$$\mathbf{HWP}(\xi,\psi) \approx \begin{bmatrix} \cos 2\psi & \sin 2\psi & -\xi\sin\psi \\ \sin 2\psi & -\cos 2\psi & \xi\cos\psi \\ \xi\sin\psi & -\xi\cos\psi & -1 \end{bmatrix}. \quad (5)$$

We now cascade a HWP with a subsequent QWP and obtain the overall rotation matrix $\mathbf{R}_{12}$ (6) (see bottom of this page). We set $\psi_i = 2\pi m_i t/T + \zeta_i$ ($i = 1, 2$) where $t$ is the time, $T$ is the depolarization interval, $\zeta_i$ are start phases or phase offsets and $m_i$ are nonzero integers. This way the waveplate eigenmodes rotate with frequencies $m_i/T$ around the equator of the Poincaré sphere.

We average $\mathbf{R}_{12}$ over $T$. To obtain the desired time average $\overline{\mathbf{R}}_{12} \approx \mathbf{0}$ (= zero matrix) in case of vanishing retardation errors, $2m_2 \neq 2m_1$ and $m_2 \neq 2m_1$ are necessary and sufficient conditions for the nonzero integers $m_i$.

If furthermore $|m_1| \neq |m_2|$ and $2m_2 \neq m_1$ holds then, to first order, the simple expression

$$\overline{\mathbf{R}}_{12} \approx \begin{bmatrix} 0 & 0 & 0 \\ 0 & 0 & 0 \\ 0 & 0 & \xi_2 \end{bmatrix} \quad (7)$$

is obtained, which depends only on the retardation error of the QWP. From (7) we obtain the largest DOP (= largest singular value of $\overline{\mathbf{R}}_{12}$) approximately as

$$DOP_{\max} \approx |\xi_2|. \quad (8)$$

Let's look at the other, alternative cases.

For $2m_2 = m_1$ we get

$$\overline{\mathbf{R}}_{12} \approx \begin{bmatrix} 0 & 0 & \xi_1/2 \cdot \sin(2\zeta_2 - \zeta_1) \\ 0 & 0 & -\xi_1/2 \cdot \cos(2\zeta_2 - \zeta_1) \\ 0 & 0 & \xi_2 \end{bmatrix} \quad (9)$$

and a corresponding, somewhat less favorable

$$DOP_{\max} \approx \sqrt{\xi_1^2/2 + \xi_2^2}. \quad (10)$$

Else, for $m_1 = -m_2$ one obtains

$$\overline{\mathbf{R}}_{12} \approx \begin{bmatrix} -\xi_1/2 \cdot \cos(\zeta_2 + \zeta_1) & -\xi_1/2 \cdot \sin(\zeta_2 + \zeta_1) & 0 \\ -\xi_1/2 \cdot \sin(\zeta_2 + \zeta_1) & \xi_1/2 \cdot \cos(\zeta_2 + \zeta_1) & 0 \\ 0 & 0 & \xi_2 \end{bmatrix} \quad (11)$$

and

$$DOP_{\max} \approx \max(|\xi_1/2|, |\xi_2|). \quad (12)$$

While this can (but need not) be worse than (8) it is all the same highly advantageous because with $m_1 = -m_2 = 1$ the waveplate driving signal period equals the depolarization interval $T$. This way $T$ can be made as short as possible, given that the needed driving frequencies are the lowest possible. This choice $m_1 = -m_2 = 1$ was also realized in the very similar case [7] for each pair of acoustooptic mode converters.

Remember that (4)–(12) are only linear approximations because we have set $\cos\xi \approx 1 - \xi^2/2 \approx 1$, thereby neglecting quadratic terms (and higher-order terms, also due to $\sin\xi \approx \xi$).

### III. Depolarizer with 3 rotating waveplates

Next, we cascade with this depolarizer yet another subsequent nonideal QWP. Using the above definitions, the rotation matrix of this sequence HWP, QWP, QWP (Fig. 1) is

$$\mathbf{R}_{123} = \mathbf{QWP}(\xi_3,\psi_3)\mathbf{QWP}(\xi_2,\psi_2)\mathbf{HWP}(\xi_1,\psi_1). \quad (13)$$

With retardation errors $\xi_i$ considered, $\mathbf{R}_{123}$ looks fairly complicated. Hence it is reasonable to analyze $\mathbf{R}_{123}$ and its time average $\overline{\mathbf{R}}_{123}$ only numerically.

$$\mathbf{R}_{12} = \mathbf{QWP}(\xi_2,\psi_2)\mathbf{HWP}(\xi_1,\psi_1) \approx$$
$$\begin{bmatrix} \begin{array}{l}(1+\xi_2)/2 \cdot \cos(2\psi_2 - 2\psi_1) \\ +(1-\xi_2)/2 \cdot \cos 2\psi_1 \\ +\xi_1/2(\cos(\psi_2 - \psi_1) - \cos(\psi_2 + \psi_1))\end{array} & \begin{array}{l}-(1+\xi_2)/2 \cdot \sin(2\psi_2 - 2\psi_1) \\ +(1-\xi_2)/2 \cdot \sin 2\psi_1 \\ -\xi_1/2(\sin(\psi_2 - \psi_1) + \sin(\psi_2 + \psi_1))\end{array} & \xi_1\begin{pmatrix}-(1-\xi_2)/2 \cdot \sin\psi_1 \\ +(1+\xi_2)/2 \cdot \sin(2\psi_2 - \psi_1)\end{pmatrix} - \sin\psi_2 \\ \begin{array}{l}(1+\xi_2)/2 \cdot \sin(2\psi_2 - 2\psi_1) \\ +(1-\xi_2)/2 \cdot \sin 2\psi_1 \\ +\xi_1/2(\sin(\psi_2 - \psi_1) - \sin(\psi_2 + \psi_1))\end{array} & \begin{array}{l}(1+\xi_2)/2 \cdot \cos(2\psi_2 - 2\psi_1) \\ -(1-\xi_2)/2 \cdot \cos 2\psi_1 \\ +\xi_1/2(\cos(\psi_2 - \psi_1) + \cos(\psi_2 + \psi_1))\end{array} & \xi_1\begin{pmatrix}(1-\xi_2)/2 \cdot \cos\psi_1 \\ -(1+\xi_2)/2 \cdot \cos(2\psi_2 - \psi_1)\end{pmatrix} + \cos\psi_2 \\ -\sin(\psi_2 - 2\psi_1) - \xi_2\xi_1\sin\psi_1 & -\cos(\psi_2 - 2\psi_1) + \xi_2\xi_1\cos\psi_1 & \xi_1\cos(\psi_2 - \psi_1) + \xi_2 \end{bmatrix} \quad (6)$$

In order to avoid accuracy loss, not the approximate matrix expressions (4), (5) but rather the exact matrices resulting from (3) were used in the numerical calculation.

For waveplate driving with

$$\psi_i = 2\pi m_i t/T + \zeta_i \qquad (i = 1, 2, 3), \tag{14}$$

combinations $[m_1, m_2, m_3]$ have been searched which result in a polarization-independent depolarizer which suffers only a low impact of the retardation errors $\xi_i$. For simplicity we set an upper limit $|\xi_i| \leq \xi_{\max}$ for all retardation errors. Exemplary suitable combinations $[m_1, m_2, m_3]$ are listed in the left part of Table 1. From top to bottom, the maximum relative driving frequency $\max|m_i|$ increases. A low $\max|m_i|$ is experimentally least complicated or permits to achieve the shortest depolarization time $T$.

For all combinations listed in Table 1,

$$DOP_{\max} \approx \xi_{\max}^2 \tag{15}$$

was found. Since the small error is squared this means a drastic relative improvement over (8), (10), (12). The result surprises because there is just a single QWP more, not a complete depolarizer (consisting of HWP and QWP) more.

The quadratic (rather than linear) dependence (13) makes it doubtful that the approximation $\cos\xi \approx 1$ in (4)–(12) would have been permissible here. This re-justifies the decision for numerical evaluation of $\mathbf{R}_{123}$, $\overline{\mathbf{R}}_{123}$.

Some observations can be or have been made. For the moment, neglect the info written in parentheses ():

- The lowest relative maximum frequency $\max|m_i|$ is 3, e.g. in $[2, 3, -2]$ ($[1, -3, 2]$).
- The lowest relative HWP frequency $|m_{HWP}|$ is 1, e.g. in $[1, 3, -3]$ ($[2, -1, 4]$).
- The lowest sum $|m_1| + |m_2| + |m_3|$ of all relative frequencies is 6, e.g. in $[1, 4, -1]$ ($[1, -3, 2]$).
- The various start phases $\zeta_i$ do not matter.
- An exchange of all frequencies by their negatives does not matter.
- The order of waveplates can be inverted, as long as each waveplate keeps its driving frequency.

So, the sequence HWP, QWP, QWP driven with $[1, 3, -3]$ is equivalent to QWP, QWP, HWP driven with $[-3, 3, 1]$ or $[3, -3, -1]$.

In addition to HWP, QWP, QWP also the sequence QWP, HWP, QWP has been investigated. This leads to

$$\mathbf{R}_{123} = \mathbf{QWP}(\xi_3, \psi_3)\mathbf{HWP}(\xi_2, \psi_2)\mathbf{QWP}(\xi_1, \psi_1). \tag{16}$$

The impact of the $\xi_i$ on $DOP_{\max}$ is somewhat different for this sequence. It is minimally higher. But the approximation (15) still roughly holds for $|\xi_i| \leq \xi_{\max}$ as long as suitable $[m_1, m_2, m_3]$ are chosen. Exemplary suitable combinations are listed in the right part of Table 1.

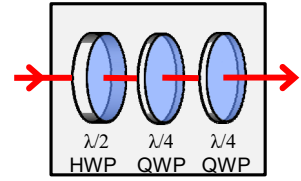

Fig. 1. Schematic representation of polarization-independent optical depolarizer, consisting of a rotating halfwave plate and two subsequent rotating quarterwave plates

TABLE 1

RELATIVE DRIVING FREQUENCIES $[m_1, m_2, m_3]$ OF ROTATING WAVEPLATES FOR TWO DIFFERENT DEPOLARIZER ARRANGEMENTS

| HWP | QWP | QWP | QWP | HWP | QWP |
|---|---|---|---|---|---|
| $m_1$ | $m_2$ | $m_3$ | $m_1$ | $m_2$ | $m_3$ |
| 1 | –3 | 3 | 1 | –3 | 2 |
| 1 | 3 | –3 | 1 | –4 | 2 |
| 2 | 3 | –2 | 1 | –4 | 3 |
| 2 | –3 | –2 | 2 | –4 | 3 |
| 3 | 1 | –3 | 2 | –1 | 4 |
| 3 | –1 | –3 | 2 | –3 | 4 |
| 3 | –2 | 2 | 3 | –2 | 4 |
| 3 | 2 | –3 | 4 | 1 | –5 |
| 3 | –2 | –3 | 1 | –5 | 2 |
| 1 | 4 | –1 | 4 | –5 | 3 |
| 4 | –1 | 1 | | | |
| 3 | –4 | –3 | | | |
| 1 | 5 | –1 | | | |
| 5 | –1 | 1 | | | |
| 4 | –5 | –4 | | | |

The observations made above (bullet points) are true also here, only for other relative frequency combinations $[m_1, m_2, m_3]$ which are given in the parentheses ().

As an alternative depolarizer one could operate $n$ independently rotating waveplates, each with retardation $\arccos(-1/3) \approx 1.91\,\mathrm{rad}$. This promises a $DOP_{\max} = 1/3^n$ [5], which appears to be less efficient than our favored solution presented here.

## IV. CONCLUSIONS

Polarization-independent optical depolarizers can be realized with a rotating halfwave and a rotating quarterwave plate. In the favorable embodiment $m_1 = -m_2 = 1$ the maximum residual degree-of-polarization is $DOP_{\max} \approx \max(|\xi_1/2|, |\xi_2|)$.

With just one more quarterwave plate, $DOP_{\max} \approx \xi_{\max}^2$ is achieved with $|\xi_i| \leq \xi_{\max} \ll 1$. This means a drastic improvement. Even though, the maximum waveplate driving frequency can be as low as 3 times the inverse of the depolarization interval $T$. This is helpful because electrooptic LiNbO$_3$ polarization transformers typically need driving voltage amplitudes of tens of Volt.